\documentclass[preprint,aps,pra,showpacs,floatfix]{revtex4-1}
\usepackage[utf8]{inputenc}
\usepackage[OT1]{fontenc}
\usepackage{amsmath}
\usepackage{amsfonts}
\usepackage{amssymb}
\usepackage{bm}
\usepackage{graphicx}
\usepackage[left=2cm,right=2cm,top=2cm,bottom=2cm]{geometry}
\usepackage{longtable}

\begin{document}

\title{QED calculation of the ground-state energy of berylliumlike ions}

\author{A. V. Malyshev$^{1}$, A. V. Volotka$^{1,2}$, D. A. Glazov$^{1,2,3}$, I. I. Tupitsyn$^{1}$,  V. M. Shabaev$^{1}$, and G. Plunien$^{2}$}

\affiliation{
$^1$ Department of Physics, St. Petersburg State University,
Ulianovskaya 1, Petrodvorets, 198504 St. Petersburg, Russia \\
$^2$ Institut f\"ur Theoretische Physik, Technische Universit\"at Dresden,
Mommsenstra{\ss}e 13, D-01062 Dresden, Germany \\
$^3$ State Scientific Centre ``Institute for Theoretical and Experimental Physics'' of National Research Centre ``Kurchatov Institute'', B. Cheremushkinskaya st. 25, 117218 Moscow, Russia \\
}

\begin{abstract}
\textit{Ab initio} QED calculations of the ground-state
binding energies of berylliumlike ions are performed for the wide range 
of the nuclear charge number: $Z=18-96$. 
 The calculations are carried out in the framework of
the extended Furry picture starting with three different types of the screening 
potential. The rigorous QED calculations up to the second order of the perturbation theory
are combined with  the third- and higher-order electron-correlation contributions
obtained within the Breit approximation by the use 
of the large-scale configuration-interaction Dirac-Fock-Sturm method. 
The effects of nuclear recoil and nuclear polarization are taken into account. 
The ionization potentials are obtained by subtracting the binding energies of the 
corresponding lithiumlike ions.
In comparison with the previous calculations the accuracy of the binding energies 
and the ionization potentials is significantly improved.
\end{abstract}

\maketitle

\section{Introduction}

%Quantum electrodynamics (QED) is well known as a very powerful tool 
%for calculation of energy levels structure in few-electrons ions. 
%This theory allows one to derive formal expressions for radiative corrections and 
%consistently take them  into account within the perturbation theory. The calculation 
%according to these formulas, however, often turns out to be a separate numerical problem. 

High precision measurements of the binding energies in highly charged ions
\cite{Schweppe:1991:1434,Stoehlker:1993:2184,Beiersdorfer:1998:1944,Beiersdorfer:1998:3022,
Bosselmann:1999:1874,Stoehlker:2000:3109,Brandau:2003:073202,Draganic:2003:183001,
Gumberidze:2004:203004,Gumberidze:2005:223001,Beiersdorfer:2005:233003,Mackel:2011:143002}
have stimulated systematic QED calculations of these systems to all orders in the nuclear strength parameter $\alpha Z$, where
$\alpha$ is the fine structure constant and $Z$ is the nuclear charge number.
To date, such calculations up to the second-order QED contributions
have been performed for H-like ions \cite{Mohr:1998:227,Yerokhin:2003:203,Yerokhin:2008:062510},
He-like ions \cite{Persson:1996:204,Yerokhin:1997:361,Artemyev:2005:062104}, Li-like ions 
  \cite{Yerokhin:2001:032109,Kozhedub:2010:042513,Sapirstein:2011:012504}, and 
B-like ions \cite{Artemyev:2007:173004,Artemyev:2013:032518}.
In other systems the QED effects were treated using either some one-electron 
approximations or semiempirical methods 
\cite{Lindroth:1992:2771,Safronova:1996:4036,Sapirstein:2002:042501,Indelicato:2005:013002,Chen:2006:042510}.
 
The main goal of the present paper is to evaluate the ground-state energies of highly charged Be-like ions 
including the QED corrections up to the second order in $\alpha$ and the electron-correlation effects
to all orders in $1/Z$.
The ground-state energies of Be-like ions
are of great importance   
%The ground state of Be-like ions is of great importance in itself, for example, 
for mass spectrometry \cite{Repp:2012:983,Myers:2013:107} and, along with investigations of H-, He-, Li-, and B-like ions,
may also serve for tests  of QED at strong fields.  
%Moreover, investigations of highly charged berylliumlike ions (along with investigations of H-, He-, Li-, and B-like systems) may serve for tests 
%of QED at strong fields. 

Calculations of the ground-state energies and 
ionization potentials of berylliumlike ions have been considered by a number of authors.
 The full-core plus correlation method was employed for the evaluation of the ionization energies for ions with $Z\leqslant 25$ in Ref.~\cite{Chung:1993:1740}. This method uses the nonrelativistic multiconfiguration 
 approach and treats the effects of relativity as the first-order perturbation.
In Ref. \cite{Chen:1997:166} the large-scale relativistic configuration-interaction 
 method with the Kohn-Sham screening potential was applied to calculate the energy levels of the ground and first excited states in ions with $Z=10-92$.  In Ref. \cite{Rodrigues:2004:117} the binding energies were calculated in the Dirac--Fock approximation for different isoelectronic series. Both binding energies and ionization potentials for ions with $Z\leqslant 60$ were evaluated in Ref.~\cite{Gu:2005:267} using the combination of the configuration interaction 
method and the many-body perturbation theory. In Ref. \cite{Huang:2006:23} the multiconfiguration Dirac-Fock method was employed and the ionization potentials for ions with $Z=37-82$ were considered. The configuration-interaction calculation
 of the energy levels of berylliumlike iron ($Z=26$) has been performed recently in Ref. \cite{Yerokhin:2014:022509}. 
All these works somehow included the radiative and nuclear recoil corrections. 
As a rule, only the first-order QED effects were incorporated. 
The necessity of a more rigorous QED treatment to improve  the theoretical accuracy has been pointed out 
 \cite{Yerokhin:2014:022509}. 

In this paper we perform \textit{ab initio} QED calculations of the ground-state
binding energies of Be-like ions in the first two orders of the perturbation theory.
The perturbative QED results are merged with the third- and higher-order electron-correlation effects
evaluated using
the large-scale configuration-interaction Dirac-Fock-Sturm method. 
The calculations are carried out for the nuclear charge number in the range: $18\leqslant Z \leqslant 96$. 
%All calculations are performed nonperturbatively in the parameter $\alpha Z$. 
%The effects of nuclear recoil in the Breit approximation are considered in all orders in $1/Z$. In addition, we take into account the QED recoil effects in the zeroth order in $1/Z$. 
In addition, we obtain the ionization potentials for Be-like ions by subtracting the binding energies of the 
corresponding lithiumlike ions. 

The paper is organized as follows. In Sec.~\ref{sec:1} we describe our approach
for  calculating the binding energies. 
In Sec.~\ref{sec:2} the numerical results for the binding energies and ionization potentials are presented. 
%The obtained results are briefly summarized in Sec.~\ref{sec:3}. 

The relativistic units ($\hbar = c = 1$) and the Heaviside charge unit ($\alpha = e^2/4\pi, e<0$) are used throughout the paper.

\section{Method of calculation \label{sec:1}}

The standard way to describe highly charged ions in the framework of QED is to use the Furry picture. 
To the zeroth order, this picture neglects the interaction between electrons and 
treats them as moving in the Coulomb field of the nucleus. 
Therefore, in the zeroth-order approximation the electrons obey the Dirac equation:
\begin{equation}
\left[-i \bm {\alpha} \cdot \nabla + \beta m + V_{\rm{nuc}}( \bm{r} )\right] \psi_n (\bm{r}) = \varepsilon_n \psi_n (\bm{r}).
\label{DirEq}
\end{equation} 
The interaction between electrons and the coupling with the quantized electromagnetic field are accounted for 
by perturbation theory. To formulate the QED perturbation theory we use 
the two-time Green function (TTGF) method \cite{TTGF}.

The convergence of the perturbation series can be accelerated
by using the extended Furry picture, that is obtained
by replacement of the nucleus potential $V_{\rm{nuc}}$ in Eq. (\ref{DirEq}) with the effective potential:
\begin{equation}
V_{\rm{nuc}}( \bm{r} ) \rightarrow V_{\rm{eff}}( \bm{r} ) = V_{\rm{nuc}}( \bm{r} ) + V_{\rm{scr}}( \bm{r} ).
\label{EffPot}
\end{equation} 
The screening part $V_{\rm{scr}}( \bm{r} )$ in Eq. (\ref{EffPot}) partly accounts for the interelectronic interaction in the zeroth-order Hamiltonian. In order to avoid the double counting the counterterm $-V_{\rm{scr}}$ must be added to the Feynman diagram technique. 
Therefore, the perturbation series are constructed in powers of the difference between the full QED interaction Hamiltonian 
and the screening potential. This approach is especially useful for low-$Z$ ions, where the interelectronic interaction becomes comparable with binding energies of all electrons involved. The extended Furry picture was successfully applied to QED calculations 
of the energy levels \cite{Sapirstein:2001:022502,Sapirstein:2002:042501,Chen:2006:042510,Artemyev:2007:173004,Yerokhin:2007:062501,Kozhedub:2010:042513,Sapirstein:2011:012504,Artemyev:2013:032518}, the $g$-factor \cite{Glazov:2006:330}, and the hyperfine splitting 
%\cite{Sapirstein:2001:032506,Sapirstein:2003:022512,Sapirstein:2006:042513,Oreshkina:2007:889:star,*Oreshkina:2007:889:transl,Kozhedub:2007:012511,Volotka:2008:062507}
\cite{Sapirstein:2001:032506,Sapirstein:2003:022512,Sapirstein:2006:042513,Oreshkina:2007:889:note,
Kozhedub:2007:012511,Volotka:2008:062507}. 
Another advantage of using the extended Furry picture, which simplifies the calculations, is 
avoiding the quasidegeneracy of the states
$1s^22s^2$ and $1s^2(2p_{1/2})^2$, that takes place 
 for the Coulomb field. 

In the present work we use three different screening potentials. 
The first choice is the local Dirac-Fock (LDF) potential which is constructed
 from the wave function of the $1s^22s^2$ state 
obtained within the Dirac-Fock approximation. 
The derivation of the  $V_{\rm{LDF}}$ potential is described in details in Ref. \cite{pot:LDF}.
% where it was successfully applied to evaluation of parity-nonconserving amplitudes in francium and cesium.
Two other potentials are derived within the density-functional theory. The Kohn-Sham screening potential can be written in a simple form, if one introduces the total radial charge density of all electrons:
\begin{gather}
\rho_t (r) = 2 \sum_{i=1s,2s} \left[ G_i^2(r) + F_i^2(r) \right] \label{rhot}\,,\\
\int_0^\infty \rho_t(r) dr = N, \label{normrhot}
\end{gather}
where $G/r$ and $F/r$ are large and small radial components of the Dirac wave functions, and $N=4$ is the total number of the electrons.
 The Kohn-Sham potential is expressed then as follows \cite{pot:KS}:
\begin{equation}
V_{\rm{KS}} = \alpha \int_0^\infty d r' \frac{1}{r_>} \rho_t(r') - \frac{2}{3} \frac{\alpha}{r} \left( \frac{81}{32\pi^2}  r \rho_t(r) \right)^{1/3}.
\label{KSpot}
\end{equation}
To improve the asymptotic behavior of this potential at large $r$ we have added the Latter correction \cite{pot:Latter}. 
%The Kohn-Sham potential was successfully utilized in QED calculations for the $g$ factor  \cite{Glazov:2006:330} and hyperfine splitting \cite{Sapirstein:2001:032506,Oreshkina:2007:815,Volotka:2008:062507}, and transition energies \cite{Sapirstein:2001:022502,Sapir2011,Kozhedub:2010:042513} in Li-like ions. 
The third potential applied in our work is  the
 Perdew-Zunger potential $V_{\rm{PZ}}$ \cite{pot:PZ}. 
%The corresponding density functional, in contrast to the Kohn-Sham density functional,
% takes into account the effects of electron-electron correlation. 
This potential has been widely employed in molecular 
and cluster calculations. 
%Along with LDF potential Perdew-Zunger potential was used in QED calculation of forbidden $2p_{3/2}-2p_{1/2}$ transition energy in Boron-like ions \cite{Artemyev:2007:173004, Artemyev:2013:032518}.

\begin{figure}
\begin{center}
\includegraphics[width=12cm]{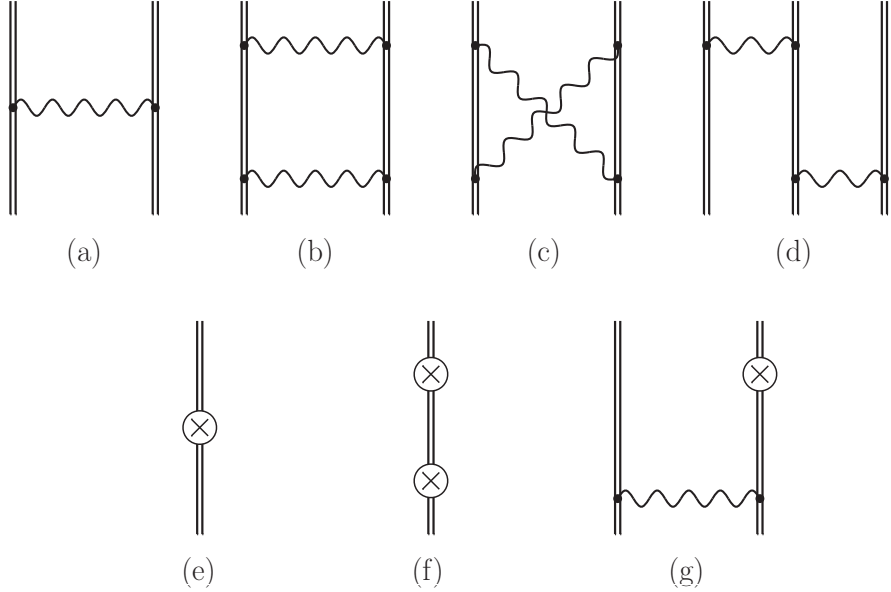}
\caption{\label{fig:int}
The interelectronic-interaction diagrams. The double line denotes the
 electron moving in the effective potential (\ref{EffPot}). The symbol $\otimes$ represents the local screening potential counterterm.}
\end{center}
\end{figure}

The calculation of the binding energies of berylliumlike ions can be divided into several stages. 
At the first step one has to solve Eq.~(\ref{DirEq}) with an effective potential. 
Moreover, to perform intermediate state summations that arise in the bound states QED calculations 
one needs to have a quasi-complete set of the Dirac equation solutions. 
The numerical evaluation of the one-electron wave functions was performed using the dual kinetic balance (DKB) 
approach \cite{splines:DKB} with the basis functions constructed from the B-splines \cite{splines:B}.

Next, we have calculated the set of Feynman diagrams describing the remaining interelectronic interaction. 
These diagrams are shown in Fig.~\ref{fig:int}. The circle with a cross denotes the screening potential counterterm. 
%These diagrams don't include any electron or photon loop and arise naturally not only in the QED approach 
%but also in relativistic non-QED  many-body calculations. 
In the most of previous works the consideration was restricted to the calculation of 
the interaction between two $1s$-electrons (He-like ions) or the interaction of $2s$-electron with 
$1s^2$ core (Li-like ions). Since in the present work we are interested in the binding energies, 
we have to calculate the diagrams depicted in Fig.~\ref{fig:int} for all possible electron configurations. 
For example, the diagram ({\it d}) is to be calculated for $1s^2 2s$ and $1s 2s^2$ subsets of electrons. 

The formulas for calculation of the diagrams ({\it a})-({\it d}) can be found, e.g., in Refs. \cite{Shabaev:1994:4489,Yerokhin:2001:032109}.
 A slight modification of the energy integration contour in the complex plane has to be performed to adopt them for the evaluation of
the $2s^2$ interaction. The derivation of the formal expressions for the ({\it e})-({\it g}) graphs within 
the TTGF method is straightforward. One easily obtains:
\begin{eqnarray}
%\Delta E_e &=&  - 2 \sum_{a=1s,2s} \langle a| V_{\rm scr} | a\rangle, \label{contr_e} \\
\Delta E_e &=&  2 \sum_{a=1s,2s} V_{aa}, \label{contr_e} \\
%%%%%%%%%%%%%%%%%%%%%%%%%
%\Delta E_f &=& 2 \sum_{a=1s,2s} \sum_{n \neq a} \frac{\langle a| V_{\rm scr} | n\rangle^2}{\varepsilon_{a} - %\varepsilon_n},   \label{contr_f} \\
\Delta E_f &=& 2 \sum_{a=1s,2s} \sum_{n \neq a} \frac{|V_{an}|^2}{\varepsilon_{a} - \varepsilon_n},   \label{contr_f} \\
%%%%%%%%%%%%%%%%%%%%%%%%%
%\Delta E_g &=& \left.-4  \sum_{a=1s,2s}  \sum_{n \neq a} \frac{I_{a \bar{a}; n \bar{a}} \langle n| V_{\rm scr} | %a\rangle}{\varepsilon_a - \varepsilon_n} \right|_{\mu_{\bar{a}}=-\mu_a} \nonumber \\
%&+& \Bigg\{ \sum_{\mu_a,\mu_b}  2  \Bigg[ 
%   \sum_{n \neq b} \frac{I_{a b ; n a} \langle n| V_{\rm scr} | b\rangle}{\varepsilon_b - \varepsilon_n}   \nonumber\\
%  &-&  \sum_{n \neq a} \frac{I_{a b ; n b}\langle n| V_{\rm scr} | a\rangle}{\varepsilon_a - \varepsilon_n}\Bigg]   %\nonumber\\
% &+& \left(\langle b| V_{\rm scr} | b\rangle - \langle a| V_{\rm scr} | a\rangle\right) I'_{baab}(\varepsilon_b - %\varepsilon_a)\Bigg\}  \Bigg|_{a=1s,b=2s} , \label{contr_g}
\Delta E_g &=& \left. 4  \sum_{a=1s,2s}  \sum_{n \neq a} \frac{I_{a \bar{a}; n \bar{a}} V_{na} }{\varepsilon_a - \varepsilon_n} \right|_{\mu_{\bar{a}}=-\mu_a} \nonumber \\
&+&  \sum_{a=1s}   \sum_{b=2s} 
\Bigg\{  2  \Bigg[ 
   \sum_{n \neq b} \frac{I_{b a ; n a} V_{nb} }{\varepsilon_b - \varepsilon_n} 
 + \sum_{n \neq a} \frac{I_{a b ; n b} V_{na} }{\varepsilon_a - \varepsilon_n} \Bigg]   \nonumber\\
 &+& \left( V_{aa} - V_{bb} \right) I'_{baab}(\varepsilon_b - \varepsilon_a)\Bigg\} 
% \Bigg|_{a=1s,b=2s} 
, \label{contr_g}
\end{eqnarray} 
where $V_{ab}=\langle a | -V_{\rm scr} | b \rangle$, $I_{abcd}(\omega) = \langle ab| I(\omega)|cd\rangle$, $I(\omega)=e^2 \alpha^\mu \alpha^\nu D_{\mu\nu}(\omega)$, $D$ is the photon propagator, $I'_{abcd}(\omega)=\langle ab| \tfrac{\partial}{\partial \omega}I(\omega)|cd\rangle$, $I_{ab;cd}=\langle ab| I(\Delta_{bd})|cd\rangle - \langle ba| I(\Delta_{ad})|cd\rangle$, 
$\Delta_{ab} = \varepsilon_a - \varepsilon_b$, and $a$ and $b$ denote the corresponding Dirac states.

\begin{figure}
\begin{center}
\includegraphics[width=13cm]{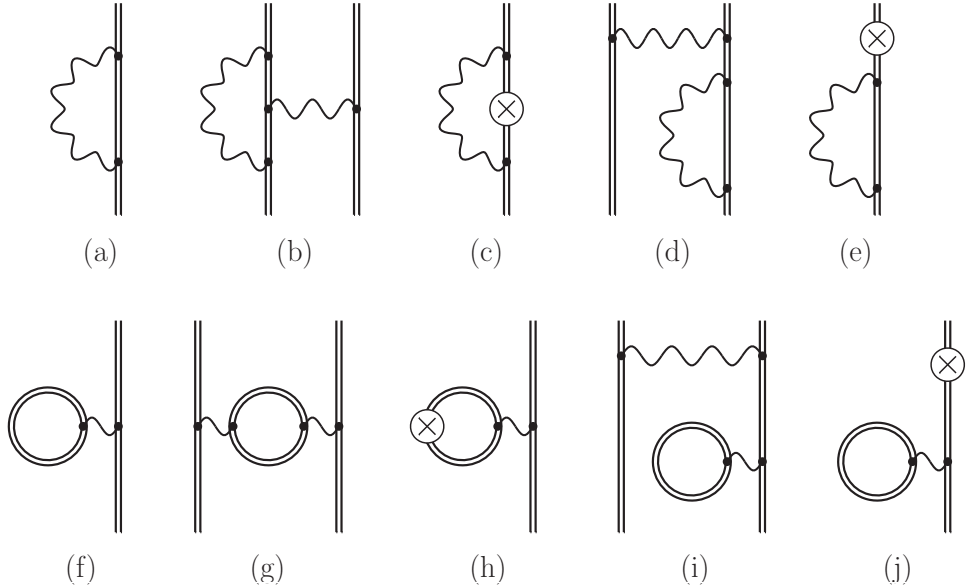}
\caption{\label{fig:se}
First- and second-order QED diagrams (excluding the one-electron two-loop diagrams). The notations are the same as in Fig. \ref{fig:int}.}
\end{center}
\end{figure}

The interelectronic-interaction contributions of the
third and higher orders are also important. 
These contributions have been calculated within the Breit approximation. 
The configuration-interaction Dirac-Fock-Sturm method (CI-DFS) \cite{CI,CI2} 
was used to solve the Dirac-Coulomb-Breit equation yielding the energy. 
The procedure of separation of the desired contribution $E_{\rm int,Breit}^{(\geqslant 3)}$ from the total result obtained in
the CI-DFS calculation was described, e.g., in Refs. \cite{Kozhedub:2010:042513,Artemyev:2013:032518}. 
At the intermediate stage of this procedure the first- and second-order interelectronic-interaction contributions, 
$E_{\rm int,Breit}^{(1)}$ and $E_{\rm int,Breit}^{(2)}$, have been extracted. To evaluate the accuracy of 
the numerical procedure we have also calculated these contributions independently using our code for the QED calculation 
but in the Breit approximation, i.e. calculating the ({\it a}), ({\it b}), ({\it d})-({\it g}) diagrams in 
Fig.~\ref{fig:int} 
in the Coulomb gauge at the zero energy transfer and neglecting
 the negative-energy contribution. 
A very good agreement between the two different approaches was found for all 
 three screening potentials.

At the next stage we should take into account the contributions from diagrams shown in Fig~\ref{fig:se}.
 In this figure, all first- and second-order QED diagrams are depicted with the exception 
of one-electron two-loop graphs, which will be discussed below. 
The diagrams in the first line are referred to as the self-energy (SE) diagrams, 
while in the second line the vacuum polarization (VP) diagrams are presented. 
Formal expressions for these diagrams derived by the TTGF method
can be found, e.g., in Ref. \cite{Kozhedub:2010:042513}. 
These expressions suffer from ultraviolet divergences.
The divergences in the SE diagrams are cancelled explicitly 
according with the renormalization scheme described in detail 
in Refs. \cite{Yerokhin:1999:3522,Yerokhin:1999:800}. 
The VP corrections are conveniently divided into the Uehling and Wichmann-Kroll terms. 
In the present work the Uehling part is calculated for all VP corrections.  
The Wichmann-Kroll contribution of the diagrams ({\it i})-({\it j}) is calculated 
by using the approximate formula for the Wichmann-Kroll potential \cite{Fainshtein:1991:559}. 
Furthermore, this approximate potential is employed for calculation of the screening correction 
to the contribution of diagram ({\it f}). The one-electron Wichmann-Kroll contributions  
in the Coulomb field are obtained using the values presented in Ref. \cite{Sapirstein:2003:042111}.  
The Wichmann-Kroll part of the diagram ({\it g}) is relatively small \cite{Artemyev:1999:45} and
has been neglected here, together with the related contribution of diagram ({\it h}).

The last second-order radiative corrections that we have to account for 
 are given by the two-loop one-electron diagrams. 
The calculation of these diagrams to all orders in $\alpha Z$
is a very complicated task which has not yet been finished. 
The latest progress in this field is related to the evaluation
of the two-loop self-energy diagrams. 
For high-$Z$ ions the calculation of 
the complete set of the two-loop self-energy  diagrams was performed in 
Ref. \cite{Yerokhin:2006:253004} 
for $n=1,2$ states ($n$ is the principal quantum number). 
For medium-$Z$ ions these corrections were calculated only for the $1s$ state 
 \cite{Yerokhin:2005:040101_R,Yerokhin:2009:040501_R}. The rest of the two-loop  contributions, 
that incorporates the diagrams with the closed fermion loop, was considered in Ref. \cite{Yerokhin:2008:062510}
(see also Ref. \cite{Mohr:1998:227} and references therein).
The so-called \textit{free-loop approximation} was employed there in case when the complete evaluation 
was not performed. In the present work, to get the two-loop corrections we interpolated the data of 
Refs. \cite{Yerokhin:2005:040101_R,Yerokhin:2006:253004,Yerokhin:2008:062510,Yerokhin:2009:040501_R}.

Next, we have to account for the nuclear recoil corrections. 
The full relativistic theory of the nuclear recoil effect  can be formulated only within QED.
Such a theory to the first order in $m/M$ ($M$ is the nuclear mass)  and
to all orders in $\alpha Z$ was developed in Refs. 
%\cite{Shabaev:1985:394:star,*Shabaev:1985:394:transl,Shabaev:1988:107:star,*Shabaev:1988:107:transl}
\cite{Shabaev:1985:394:note,Shabaev:1988:107:note}  
(see also Refs. \cite{Shabaev:1998:59,Adkins:2007:042508} and references therein). 
In the Breit approximation the theory leads to the following Hamiltonian
\cite{Shabaev:1985:394:note,Shabaev:1988:107:note,Palmer:1987:5987}:
\begin{equation}
H_{M}=\frac{1}{2M} \sum_{i,j} \left\{ \bm{p}_i \cdot \bm{p}_j  - \frac{\alpha Z}{r_i} \left[ \bm{\alpha}_i + \frac{(\bm{\alpha}_i \cdot \bm{r}_i) \bm{r}_i }{r_i^2} \right] \cdot \bm{p}_j \right\}.
\label{BrRecoil}
\end{equation}
To get the nuclear recoil correction within the Breit approximation we evaluated
the expectation value of the Hamiltonian (\ref{BrRecoil})
with the wave functions obtained by the CI-DFS method. 
The nuclear recoil corrections beyond the Breit approximation, 
which are referred to as the QED nuclear recoil effects, 
have been evaluated to the zeroth order in $1/Z$. 
Since all the electrons in the ground state of a berylliumlike ion ($1s^22s^2$) have the same parity, 
in the zeroth order in $1/Z$ 
the two-electron QED recoil corrections vanish 
\cite{Artemyev:1995:1884}. Therefore, we have to account for the one-electron QED recoil corrections 
only and, therefore, can use the related results for hydrogenlike ions. 
In Refs. \cite{Shabaev:1998:4235,Shabaev:1999:493} the one-electron QED recoil corrections have been 
calculated for extended nuclei to all orders in $\alpha Z$. Here we interpolate the data from these works.

Finally, for high-$Z$ ions one has to take into account the effect of nuclear polarization.
This correction results from the electron-nucleus interaction
diagrams, in which the intermediate nuclear states are excited. 
We incorporate this correction using the results of Refs. 
%\cite{Plunien:1995:1119:star,*Plunien:1995:1119:erratum,Nefiodov:1996:227,Volotka:2014:023002}.
\cite{Plunien:1995:1119:note,Nefiodov:1996:227,Volotka:2014:023002}. 

To complete the discussion of the computation details it is worth noting that 
the numerical procedure for the evaluation of the QED corrections was checked by using two gauges, 
the Feynman and the Coulomb ones. Both calculations agreed very well with each other. The calculations 
were performed for extended nuclei using the Fermi nuclear-charge
distribution with a thickness parameter of $2.3$~fm. 
The nuclear radii were taken from Refs. \cite{Angeli:2013:69}.

\section{Numerical results and discussions \label{sec:2}}

In this section we present our results for the ground-state
binding energies and ionization potentials in berylliumlike ions. 

{
\renewcommand{\arraystretch}{0.85}
\begin{table}
\caption{Individual contributions to the ground-state binding energy of berylliumlike calcium (in eV). 
%Nucleus radius is 3.4776(19)~fm. 
See text for details.}\label{table:Ca}
\begin{tabular}{c@{\quad}r@{\quad}r@{\quad}r}
\hline 
\hline 
\multicolumn{1}{c}{Contribution} & \multicolumn{1}{c}{LDF} & \multicolumn{1}{c}{KS} & \multicolumn{1}{c}{PZ}     \\ 
\hline 

$E^{(0)}_{\rm Dirac}$              &  $-11618.844$    &  $-11704.702$    &  $-11782.120$     \\ 

$E^{(1)}_{\rm int}$                &  $-1219.511$     &  $-1134.281$     &  $-1056.673$      \\ 

$E^{(2)}_{\rm int,Breit}$          &  $-10.996$       &  $-9.725$        &  $-11.997$        \\ 

$E^{(2)}_{\rm int,QED}$            &  $0.007$         &  $0.007$         &  $0.007$          \\ 

$E^{(\geqslant3)}_{\rm int,Breit}$ &  $1.983$         &  $1.340$         &  $3.423$          \\ 

\hline 
$E_{\rm int,total}$                & $-12847.360$     & $-12847.360$     & $-12847.359$      \\ 
\hline 

$E^{(1)}_{\rm QED}$                &  $3.366$         &  $3.419$         &  $3.383$          \\ 

$E^{(2)}_{\rm ScrQED}$             &  $0.086$         &  $0.032$         &  $0.068$          \\ 

$E^{(2l)}_{\rm QED}$               &  $-0.003$        &  $-0.003$        &  $-0.003$         \\ 

\hline 
$E_{\rm QED,total}$                &  $3.449$         &  $3.448$         &  $3.448$          \\ 
\hline 
  
$E_{\rm Rec,Breit}$                &  $0.176$         &  $0.176$         &  $0.176$          \\ 

$E_{\rm Rec,QED}$                  &  $0.001$         &  $0.001$         &  $0.001$          \\ 

\hline 
\hline 
$E_{\rm total}$                    & $-12843.734$     & $-12843.736$     & $-12843.735$      \\ 
\hline 
\hline 
\end{tabular} 
\end{table}
}

The individual contributions 
to the ground-state
 binding energies of berylliumlike calcium, xenon, and uranium calculated for the three screening potentials are collected in Tables \ref{table:Ca}--\ref{table:U}, respectively.  
%For discussion of the results it is convenient also to present the particular terms calculated in the pure Coulomb potential. As it was mentioned above to perform the full calculation in the Coulomb potential without the reconstruction of the perturbation theory appears to be impossible due to the quasi-degeneracy of $2s^2$ and $2p^2_{1/2}$ states.
For each ion the first line displays the binding energy $E^{(0)}_{\rm Dirac}$
obtained as a sum of the one-electron Dirac energies. For uranium we added the nuclear deformation correction following the results of Ref.~\cite{Kozhedub:2008:032501}.
In the second line the contribution of the first-order diagrams presented in
Fig.~\ref{fig:int} is given (diagrams ({\it a}) and ({\it e})). The diagram ({\it a}) is calculated within QED, i.e. 
the difference of the reference-state energies is kept in the photon propagator of the exchange part. 
In the third row we give the contribution of the second-order diagrams in 
Fig.~\ref{fig:int} evaluated using the Breit approximation. We note 
that we consider the Coulomb and Breit photons on an equal footing. Therefore, the exchange by two Breit photons belongs to the second order term. This way 
to account for the Breit interaction differs from the way of Yerokhin \textit{et al.} \cite{Yerokhin:2007:062501}, where the Breit interaction is considered to the first order only. Moreover, in the Ref. \cite{Yerokhin:2007:062501} the negative energy continuum was partly accounted for in the correction under consideration. 
The fourth line contains the QED correction $E^{(2)}_{\rm int,QED}$ to the third line, that is defined as
the difference between the calculations of the second-order diagrams in the framework of the rigorous QED approach
and the Breit approximation. In the fifth line  we give the electron-correlation contribution 
 of the third and higher orders  in the Breit approximation obtained from the CI-DFS calculations. 
The sixth row displays the  sum  of  all the previous terms (lines from first to fifth), $E_{\rm int,total}$. 
From Tables \ref{table:Ca}--\ref{table:U} it is seen that the 
$E_{\rm int,total}$ values are in a good agreement with each other for all screening potentials. 
%Since the total CI-DFS energies are independent of the effective potential one can start from, it means mainly a good agreement of the sums of the QED corrections to the first and second orders.
%For the pure Coulomb potential we didn't calculate the second order of the interelectronic interaction. Instead, we extracted from the CI-DFS results the sum  of the second- and higher-orders in the Breit approximation, this quantity is placed in the fifth line. The uncalculated QED correction was roughly estimated  by multiplying  of the Breit contribution by the factor $2(\alpha Z)^3$. The resulting uncertainty is given in the sixth line. The sum of all contributions for pure Coulomb potential is close to the values obtained for the screening potentials, however the estimated uncertainty is significantly greater.
{
\renewcommand{\arraystretch}{0.85}
\begin{table}
\caption{Individual contributions to the ground-state binding energy of berylliumlike xenon (in eV). 
%Nucleus radius is 4.7859(48)~fm. 
See text for details.}\label{table:Xe}
\begin{tabular}{c@{\quad}r@{\quad}r@{\quad}r}
\hline 
\hline 
\multicolumn{1}{c}{Contribution} & \multicolumn{1}{c}{LDF} & \multicolumn{1}{c}{KS} & \multicolumn{1}{c}{PZ}     \\ 
\hline 

$E^{(0)}_{\rm Dirac}$              &  $-97572.523$   &  $-97768.436$   &  $-98000.517$    \\ 

$E^{(1)}_{\rm int}$                &  $-3485.820$    &  $-3290.179$    &  $-3058.046$     \\ 

$E^{(2)}_{\rm int,Breit}$          &  $-15.206$      &  $-14.299$      &  $-16.762$       \\ 

$E^{(2)}_{\rm int,QED}$            &  $0.264$        &  $0.248$        &  $0.265$         \\ 

$E^{(\geqslant3)}_{\rm int,Breit}$ &  $1.745$        &  $1.127$        &  $3.521$         \\ 

\hline 
$E_{\rm int,total}$                &  $-101071.540$  &  $-101071.538$  &  $-101071.540$   \\ 
\hline 

$E^{(1)}_{\rm QED}$                &  $97.763$       &  $98.240$       &  $97.904$        \\ 

$E^{(2)}_{\rm ScrQED}$             &  $0.651$        &  $0.170$        &  $0.507$         \\ 

$E^{(2l)}_{\rm QED}$               &  $-0.242$       &  $-0.242$       &  $-0.242$        \\ 

\hline 
$E_{\rm QED,total}$                &  $98.173$       &  $98.168$       &  $98.169$        \\ 
\hline 
  
$E_{\rm Rec,Breit}$                &  $0.404$        &  $0.404$        &  $0.404$         \\ 

$E_{\rm Rec,QED}$                  &  $0.045$        &  $0.045$        &  $0.045$         \\ 

\hline 
\hline 
$E_{\rm total}$                    &  $-100972.919$  &  $-100972.921$  &  $-100972.922$    \\ 
\hline 
\hline 
\end{tabular} 
\end{table} 
}
{
\renewcommand{\arraystretch}{0.85}
\begin{table}
\caption{Individual contributions to the ground-state
binding energy of berylliumlike uranium (in eV). 
%Nucleus radius is 5.8571(33)~fm. 
See text for details.}\label{table:U}
\begin{tabular}{c@{\quad}r@{\quad}r@{\quad}r}
\hline 
\hline 
\multicolumn{1}{c}{Contribution} & \multicolumn{1}{c}{LDF} & \multicolumn{1}{c}{KS} & \multicolumn{1}{c}{PZ}     \\ 
\hline 

$E^{(0)}_{\rm Dirac}$              &  $-320572.56$   &  $-320871.02$   &  $-321276.02$     \\ 

$E^{(1)}_{\rm int}$                &  $-6637.37$     &  $-6338.62$     &  $-5933.84$       \\ 

$E^{(2)}_{\rm int,Breit}$          &  $-20.88$       &  $-20.45$       &  $-23.97$         \\ 

$E^{(2)}_{\rm int,QED}$            &  $2.04$         &  $1.96$         &  $2.05$           \\ 

$E^{(\geqslant3)}_{\rm int,Breit}$ &  $3.03$         &  $2.39$         &  $6.05$           \\ 

\hline 
$E_{\rm int,total}$                &  $-327225.74$   &  $-327225.74$   &  $-327225.74$     \\ 
\hline 

$E^{(1)}_{\rm QED}$                &  $618.61$       &  $620.18$       &  $618.97$         \\ 

$E^{(2)}_{\rm ScrQED}$             &  $0.73$         &  $-0.86$        &  $0.36$           \\ 

$E^{(2l)}_{\rm QED}$               &  $-2.92$        &  $-2.92$        &  $-2.92$          \\ 
 
\hline 
$E_{\rm QED,total}$                &  $616.41$       &  $616.40$       &  $616.41$         \\ 
\hline 
  
$E_{\rm Rec,Breit}$                &  $0.60$         &  $0.60$         &  $0.60$           \\ 

$E_{\rm Rec,QED}$                  &  $0.56$         &  $0.56$         &  $0.56$           \\ 

$E_{\rm Nucl.Pol.}$                &  $-0.45$        &  $-0.45$        &  $-0.45$          \\

\hline 
\hline 
$E_{\rm total}$                    &  $-326608.62$   &  $-326608.63$   &  $-326608.63$     \\ 
\hline 
\hline 
\end{tabular} 
\end{table}
}
The contributions of the first- and second-order diagrams in Fig.~\ref{fig:se} are given in the seventh and eighth rows, respectively. The ninth line contains the contribution of the two-loop one-electron diagrams.  In the row labeled $E_{\rm QED,total}$ we present the  sum of the contributions of the QED diagrams (lines from seventh to ninth). Again one can see that 
the results of the calculations for all screening potentials are in a good agreement with each other.
In the next two lines  we give the nuclear recoil correction calculated in the Breit approximation using the CI-DFS method and the QED recoil effect evaluated to the zeroth order in $1/Z$, respectively. In Table~\ref{table:U} for berylliumlike uranium 
 the row $E_{\rm Nucl.Pol.}$  presents the contribution of the nuclear polarization effect.
Finally, the total values of the ground-state
binding energies are given in the last lines.  From Tables~\ref{table:Ca}--\ref{table:U} it is seen
 that the total values of the binding energies
are almost independent of the screening potential. Hence, 
for all other ions we have performed the calculations using only the LDF screening potential.

{
\renewcommand{\arraystretch}{0.85}
\begin{table}
\footnotesize
\caption{Ground-state binding energies (in eV) of berylliumlike ions with $Z=18-96$}\label{table:all}

\begin{tabular}{llll@{\qquad}llll}
\hline 
\hline 
 \multicolumn{1}{c}{Nucl.}& \multicolumn{1}{c}{This work}& \multicolumn{1}{c}{Other theory}& \multicolumn{1}{c}{NIST} &
 \multicolumn{1}{c}{Nucl.}& \multicolumn{1}{c}{This work}& \multicolumn{1}{c}{Other theory}& \multicolumn{1}{c}{NIST} \\ 
\hline 
 
  $^{40}_{18}$Ar   &   $-10321.030(40)$   &   $-10321.23^{\rm a}$    &  $-10320.73(30)$   &
     
          $^{132}_{54}$Xe   &   $-100972.921(85)$   &   $-100973.7^{\rm a}$   &   $-100963(4)$   \\
          
  $^{40}_{20}$Ca   &   $-12843.735(41)$   &   $-12843.96^{\rm a}$    &  $-12843.29(40)$   &
     
                            &                       &   $-100973.75^{\rm b}$  &                  \\
           
                   &                      &   $-12843.989^{\rm b}$   &     &
     
          $^{138}_{56}$Ba   &  $-109050.461(95)$    &  $-109051.1^{\rm a}$    &  $-109039(4)$    \\
          
  $^{48}_{22}$Ti   &  $-15646.995(43)$    &   $-15647.31^{\rm a}$   &  $-15646.42(50)$    &
     
          $^{140}_{58}$Ce   &  $-117486.45(11)$     &  $-117486.8^{\rm a}$    &  $-117476(5)$    \\
          
  $^{52}_{24}$Cr   &  $-18732.687(44)$    &   $-18733.04^{\rm a}$   &  $-18731.96(50)$    &
     
          $^{142}_{60}$Nd   &  $-126288.97(12)$     &   $-126288.9^{\rm a}$   &  $-126279(5)$    \\
          
  $^{56}_{26}$Fe   &  $-22102.960(45)$    &   $-22103.37^{\rm a}$   &  $-22102.1(1.8)$    & 
     
          $^{152}_{62}$Sm   &  $-135465.89(14)$     &                         &  $-135455(6)$    \\
          
                   &                      &   $-22103.299^{\rm b}$  &                     & 
     
          $^{158}_{64}$Gd   &  $-145027.69(16)$     &   $-145028.63^{\rm b}$  &  $-145017(10)$   \\
          
                   &                      &  $-22102.98(8)^{\rm c}$ &                     & 
     
          $^{164}_{66}$Dy   &  $-154983.99(20)$     &                         &  $-154974(10)$   \\
           
  $^{58}_{28}$Ni   &  $-25760.181(45)$    &  $-25760.64^{\rm a}$    &  $-25759.1(2.1)$    & 
     
          $^{166}_{68}$Er   &  $-165345.88(21)$     &                         &  $-165336(20)$   \\
          
  $^{64}_{30}$Zn   &  $-29706.959(46)$    &  $-29707.48^{\rm a}$    &  $-29705.7(2.5)$    & 
     
          $^{174}_{70}$Yb   &  $-176124.31(25)$     &                         &  $-176110(40)$   \\
          
  $^{74}_{32}$Ge   &  $-33946.148(47)$    &  $-33946.75^{\rm a}$    &  $-33945(3)$        & 
     
          $^{180}_{72}$Hf   &  $-187332.39(28)$     &                         &  $-187320(50)$   \\
           
                   &                      &  $-33946.575^{\rm b}$   &                     & 
     
          $^{184}_{74}$W    &  $-198983.71(32)$     &   $-198984.71^{\rm b}$  &  $-198987(3)$    \\
          
  $^{80}_{34}$Se   &  $-38480.835(47)$    &  $-38481.51^{\rm a}$    &  $-38479(3)$        & 
     
          $^{192}_{76}$Os   &  $-211091.88(37)$     &                         &  $-211080(70)$   \\
          
  $^{84}_{36}$Kr   &  $-43314.394(48)$    &  $-43315.13^{\rm a}$    &  $-43313(4)$        & 
     
          $^{194}_{78}$Pt   &  $-223674.28(44)$     &                         &  $-223660(90)$   \\
          
  $^{88}_{38}$Sr   &  $-48450.485(50)$    &  $-48451.27^{\rm a}$    &  $-48449(4)$        & 
     
          $^{202}_{80}$Hg   &  $-236746.09(52)$     &                         &  $-236730(100)$  \\
          
  $^{90}_{40}$Zr   &  $-53893.040(51)$    &  $-53893.88^{\rm a}$    &  $-53891(5)$        & 
     
          $^{208}_{82}$Pb   &  $-250326.80(59)$     &   $-250327.64^{\rm b}$  &  $-250310(100)$  \\
          
  $^{98}_{42}$Mo   &  $-59646.278(54)$    &  $-59647.22^{\rm a}$    &  $-59644(5)$        & 
     
          $^{210}_{84}$Po   &  $-264434.5(1.2)$     &                         &  $-264430(100)$  \\
          
                   &                      &  $-59646.894^{\rm b}$   &                     & 
     
          $^{220}_{86}$Rn   &  $-279089.7(1.5)$     &                         &  $-279070(200)$  \\
          
  $^{102}_{44}$Ru  &  $-65714.838(57)$    &  $-65715.84^{\rm a}$    &  $-65712(6)$        & 
     
          $^{226}_{88}$Ra   &  $-294321.8(2.5)$     &                         &  $-294310(200)$  \\
          
  $^{106}_{46}$Pd  &  $-72103.637(61)$    &  $-72104.68^{\rm a}$    &  $-72100(6)$        &
     
          $^{232}_{90}$Th   &  $-310152.0(1.7)$     &  $-310152.4^{\rm b}$    &  $-310140(200)$  \\
          
  $^{114}_{48}$Cd  &  $-78817.939(65)$    &  $-78819.01^{\rm a}$    &  $-78814(7)$        &
     
          $^{238}_{92}$U    &  $-326608.6(1.3)$     &  $-326608.5^{\rm b}$    &  $-326600(300)$  \\
          
  $^{120}_{50}$Sn  &  $-85863.480(70)$    &  $-85864.48^{\rm a}$    &  $-85860(7)$        &
     
          $^{240}_{94}$Pu   &  $-343730.0(5.5)$     &                         &  $-343700(300)$  \\
          
  $^{130}_{52}$Te  &  $-93246.244(77)$    & $-93247.17^{\rm a}$     &  $-93236(4)$        &
     
          $^{244}_{96}$Cm   &  $-361552.4(3.6)$     &                         &  $-361500(400)$  \\        
 
\hline
\hline
\end{tabular}  

\bigskip
\raggedright

$^{\rm a}$ Gu \cite{Gu:2005:267}.

$^{\rm b}$ Chen and Cheng \cite{Chen:1997:166}  with the finite nuclear size correction 
recalculated employing the nuclear charge radii taken from Ref.~\cite{Angeli:2013:69}.

$^{\rm c}$ Yerokhin \textit{et al.} \cite{Yerokhin:2014:022509}.
\end{table}
}

The binding energy of an ion can be obtained by summing
 the ionization energies of all its constituent electrons. In case of a berylliumlike ion 
it means that one should sum the ionization potentials of the  H-, He-, Li-, and Be-like ions. 
The corresponding compilation can be found in the NIST database \cite{NIST:2013}. In this compilation 
the ionization energies of  hydrogenlike ions are taken from the work of Johnson and Soff \cite{Johnson:1985:405}. The ionization potentials for heliumlike ions almost for all nuclei of interest are obtained from the paper by Artemyev \textit{et al.} \cite{Artemyev:2005:062104}. The exceptions are the He-like radium, thorium, plutonium and curium. For these ions the energies are taken from the work of Drake \cite{Drake:1988:586}. The ionization potentials for lithiumlike ions were calculated by Sapirstein and Cheng \cite{Sapirstein:2011:012504}. For ionization energies of berylliumlike ions with $Z \leqslant 50$ the NIST database uses the tabulation by Bi\'emont \textit{et al.} \cite{Biemont:1999:117}. For tungsten ($Z=74$) the energy is obtained from the paper by Kramida and Reader \cite{Kramida:2006:457}. For all other berylliumlike ions the potentials are extracted from the work of Rodrigues \textit{et al.} \cite{Rodrigues:2004:117}. The works \cite{Biemont:1999:117} and \cite{Kramida:2006:457} are closely related to experiment, since the data presented there were obtained by combining the theoretical calculations with the systematic consideration of available spectroscopic data along different isoelectronic series.

{
\renewcommand{\arraystretch}{0.8}
\begin{table}
\begin{center}
\footnotesize
\caption{Ionization potentials (in eV) for berylliumlike ions with $Z=18-96$}\label{table:IP}

\begin{tabular}{ll@{\quad}l@{\qquad\quad}ll@{\quad}l}
\hline 
\hline 
 \multicolumn{1}{c}{Nucleus} & \multicolumn{1}{c}{This work} & \multicolumn{1}{c}{Other work} & 
 \multicolumn{1}{c}{Nucleus} & \multicolumn{1}{c}{This work} & \multicolumn{1}{c}{Other work} \\ 
\hline 
%%%%%%%%%%%%%%%%%%%%%%%%%%%%%%%%%%%%%%%%%%%%%%%%%%%%%%%%%%%%%%%%%%%%%%%%%%%%%%%%%%%%%%%%%%%%%%%%%%%%%%%%
  $^{40}_{18}$Ar   &   $-855.754(39)$    &   $-855.47(27)^{\rm a}$    &
  
				       $^{120}_{50}$Sn   &   $-8106.856(64)$          &  $-8103.1(7.3)^{\rm a}$    \\
%%%%%%%%%%%%%%%%%%%%%%%%%%%%%%%%%%%%%%%%%%%%%%%%%%%%%%%%%%%%%%%%%%%%%%%%%%%%%%%%%%%%%%%%%%%%%%%%%%%%%%%%				       
                   &                     &   $-855.82^{\rm b}$        &
       
                                         &                            &  $-8107.2^{\rm b}$         \\ 
%%%%%%%%%%%%%%%%%%%%%%%%%%%%%%%%%%%%%%%%%%%%%%%%%%%%%%%%%%%%%%%%%%%%%%%%%%%%%%%%%%%%%%%%%%%%%%%%%%%%%%%%                                                
  $^{40}_{20}$Ca   &   $-1087.273(41)$   &  $-1086.85(40)^{\rm a}$    &
  
  					   $^{130}_{52}$Te   &  $-8831.121(70)$           &  $-8821(4)^{\rm d}$        \\
%%%%%%%%%%%%%%%%%%%%%%%%%%%%%%%%%%%%%%%%%%%%%%%%%%%%%%%%%%%%%%%%%%%%%%%%%%%%%%%%%%%%%%%%%%%%%%%%%%%%%%%%				     
                   &                     &  $-1087.3^{\rm b}$         &
       
                                         &                            &  $-8831.5^{\rm b}$         \\
%%%%%%%%%%%%%%%%%%%%%%%%%%%%%%%%%%%%%%%%%%%%%%%%%%%%%%%%%%%%%%%%%%%%%%%%%%%%%%%%%%%%%%%%%%%%%%%%%%%%%%%%                                            
                   &                     &  $-1087.311^{\rm c}$       &
              
       				   $^{132}_{54}$Xe   &   $-9590.905(76)$          &  $-9581(4)^{\rm d}$        \\
%%%%%%%%%%%%%%%%%%%%%%%%%%%%%%%%%%%%%%%%%%%%%%%%%%%%%%%%%%%%%%%%%%%%%%%%%%%%%%%%%%%%%%%%%%%%%%%%%%%%%%%%  	       
  $^{48}_{22}$Ti   &   $-1346.891(43)$   &  $-1346.33(47)^{\rm a}$    &                  
  
 										 &                            &  $-9591.4^{\rm b}$         \\
%%%%%%%%%%%%%%%%%%%%%%%%%%%%%%%%%%%%%%%%%%%%%%%%%%%%%%%%%%%%%%%%%%%%%%%%%%%%%%%%%%%%%%%%%%%%%%%%%%%%%%%%				 
                   &                     &  $-1347.0^{\rm b}$         &
       
                       $^{138}_{56}$Ba   &  $-10387.024(82)$          &  $-10376(4)^{\rm d}$       \\
%%%%%%%%%%%%%%%%%%%%%%%%%%%%%%%%%%%%%%%%%%%%%%%%%%%%%%%%%%%%%%%%%%%%%%%%%%%%%%%%%%%%%%%%%%%%%%%%%%%%%%%%       
  $^{52}_{24}$Cr   &   $-1634.822(44)$   &  $-1634.11(55)^{\rm a}$    &       
 
                                         &                            &  $-10388^{\rm b}$          \\
%%%%%%%%%%%%%%%%%%%%%%%%%%%%%%%%%%%%%%%%%%%%%%%%%%%%%%%%%%%%%%%%%%%%%%%%%%%%%%%%%%%%%%%%%%%%%%%%%%%%%%%% 
                   &                     &  $-1634.9^{\rm b}$         &
       
                       $^{140}_{58}$Ce   &  $-11220.379(91)$          &  $-11210(5)^{\rm d}$       \\
%%%%%%%%%%%%%%%%%%%%%%%%%%%%%%%%%%%%%%%%%%%%%%%%%%%%%%%%%%%%%%%%%%%%%%%%%%%%%%%%%%%%%%%%%%%%%%%%%%%%%%%%       
  $^{56}_{26}$Fe   &   $-1951.307(44)$   &   $-1950.4(1.8)^{\rm a}$   &
  
                                         &                            &  $-11221^{\rm b}$          \\
%%%%%%%%%%%%%%%%%%%%%%%%%%%%%%%%%%%%%%%%%%%%%%%%%%%%%%%%%%%%%%%%%%%%%%%%%%%%%%%%%%%%%%%%%%%%%%%%%%%%%%%%  
                   &                     &   $-1951.4^{\rm b}$        &    
       
                       $^{142}_{60}$Nd   &   $-12091.91(10)$          &  $-12082(5)^{\rm d}$       \\
%%%%%%%%%%%%%%%%%%%%%%%%%%%%%%%%%%%%%%%%%%%%%%%%%%%%%%%%%%%%%%%%%%%%%%%%%%%%%%%%%%%%%%%%%%%%%%%%%%%%%%%%       
  $^{58}_{28}$Ni   &   $-2296.621(45)$   &   $-2295.6(2.1)^{\rm a}$   &
  
                                         &                            &  $-12092^{\rm b}$          \\
%%%%%%%%%%%%%%%%%%%%%%%%%%%%%%%%%%%%%%%%%%%%%%%%%%%%%%%%%%%%%%%%%%%%%%%%%%%%%%%%%%%%%%%%%%%%%%%%%%%%%%%%  
                   &                     &   $-2296.7^{\rm b}$        &         
       
                       $^{152}_{62}$Sm   &   $-13002.63(11)$          &  $-12992(6)^{\rm d}$       \\
%%%%%%%%%%%%%%%%%%%%%%%%%%%%%%%%%%%%%%%%%%%%%%%%%%%%%%%%%%%%%%%%%%%%%%%%%%%%%%%%%%%%%%%%%%%%%%%%%%%%%%%%       
  $^{64}_{30}$Zn   &   $-2671.061(46)$   &  $-2669.9(2.5)^{\rm a}$    &
  
                       $^{158}_{64}$Gd   &   $-13953.70(12)$          &  $-13943(10)^{\rm d}$      \\
%%%%%%%%%%%%%%%%%%%%%%%%%%%%%%%%%%%%%%%%%%%%%%%%%%%%%%%%%%%%%%%%%%%%%%%%%%%%%%%%%%%%%%%%%%%%%%%%%%%%%%%%  
                   &                     &  $-2671.2^{\rm b}$         &
       
                       $^{164}_{66}$Dy   &   $-14946.29(14)$          &  $-14936(15)^{\rm d}$      \\
%%%%%%%%%%%%%%%%%%%%%%%%%%%%%%%%%%%%%%%%%%%%%%%%%%%%%%%%%%%%%%%%%%%%%%%%%%%%%%%%%%%%%%%%%%%%%%%%%%%%%%%%       
  $^{74}_{32}$Ge   &   $-3074.968(46)$   &  $-3073.6(3.0)^{\rm a}$    &
  
                       $^{166}_{68}$Er   &   $-15981.70(16)$          &  $-15971(24)^{\rm d}$      \\
%%%%%%%%%%%%%%%%%%%%%%%%%%%%%%%%%%%%%%%%%%%%%%%%%%%%%%%%%%%%%%%%%%%%%%%%%%%%%%%%%%%%%%%%%%%%%%%%%%%%%%%%  
                   &                     &  $-3075.1^{\rm b}$         &
       
                       $^{174}_{70}$Yb   &   $-17061.27(17)$          &  $-17050(40)^{\rm d}$      \\
%%%%%%%%%%%%%%%%%%%%%%%%%%%%%%%%%%%%%%%%%%%%%%%%%%%%%%%%%%%%%%%%%%%%%%%%%%%%%%%%%%%%%%%%%%%%%%%%%%%%%%%%       
  $^{80}_{34}$Se   &   $-3508.689(47)$   &  $-3507.1(3.5)^{\rm a}$    &
  
                       $^{180}_{72}$Hf   &   $-18186.56(20)$          &  $-18176(50)^{\rm d}$      \\
%%%%%%%%%%%%%%%%%%%%%%%%%%%%%%%%%%%%%%%%%%%%%%%%%%%%%%%%%%%%%%%%%%%%%%%%%%%%%%%%%%%%%%%%%%%%%%%%%%%%%%%%  
                   &                     &  $-3508.9^{\rm b}$         &
       
                       $^{184}_{74}$W    &   $-19359.16(22)$          &   $-19348(50)^{\rm d}$     \\
%%%%%%%%%%%%%%%%%%%%%%%%%%%%%%%%%%%%%%%%%%%%%%%%%%%%%%%%%%%%%%%%%%%%%%%%%%%%%%%%%%%%%%%%%%%%%%%%%%%%%%%%       
  $^{84}_{36}$Kr   &   $-3972.628(47)$   &  $-3970.8(4.0)^{\rm a}$    &       

                                         &                            &   $-19362.5(3.1)^{\rm e}$  \\
%%%%%%%%%%%%%%%%%%%%%%%%%%%%%%%%%%%%%%%%%%%%%%%%%%%%%%%%%%%%%%%%%%%%%%%%%%%%%%%%%%%%%%%%%%%%%%%%%%%%%%%%
                   &                     &  $-3972.8^{\rm b}$         &
       
                       $^{192}_{76}$Os   &   $-20580.79(25)$          &   $-20570(70)^{\rm d}$     \\
%%%%%%%%%%%%%%%%%%%%%%%%%%%%%%%%%%%%%%%%%%%%%%%%%%%%%%%%%%%%%%%%%%%%%%%%%%%%%%%%%%%%%%%%%%%%%%%%%%%%%%%%       
  $^{88}_{38}$Sr   &   $-4467.212(48)$   &  $-4465.3(4.5)^{\rm a}$    &                                                     
  
                       $^{194}_{78}$Pt   &   $-21853.51(28)$          &   $-21843(90)^{\rm d}$     \\
%%%%%%%%%%%%%%%%%%%%%%%%%%%%%%%%%%%%%%%%%%%%%%%%%%%%%%%%%%%%%%%%%%%%%%%%%%%%%%%%%%%%%%%%%%%%%%%%%%%%%%%%  
                   &                     &  $-4467.4^{\rm b}$         & 
       
                       $^{202}_{80}$Hg   &   $-23179.15(31)$          &   $-23168(110)^{\rm d}$    \\       
%%%%%%%%%%%%%%%%%%%%%%%%%%%%%%%%%%%%%%%%%%%%%%%%%%%%%%%%%%%%%%%%%%%%%%%%%%%%%%%%%%%%%%%%%%%%%%%%%%%%%%%%       
  $^{90}_{40}$Zr   &   $-4992.900(50)$   &  $-4990.7(5.0)^{\rm a}$    &    
  
                       $^{208}_{82}$Pb   &   $-24560.05(34)$          &   $-24548(120)^{\rm d}$    \\
%%%%%%%%%%%%%%%%%%%%%%%%%%%%%%%%%%%%%%%%%%%%%%%%%%%%%%%%%%%%%%%%%%%%%%%%%%%%%%%%%%%%%%%%%%%%%%%%%%%%%%%%  
                   &                     &  $-4993.1^{\rm b}$         &  
       
                       $^{210}_{84}$Po   &   $-25998.65(42)$          &   $-25988(150)^{\rm d}$    \\
%%%%%%%%%%%%%%%%%%%%%%%%%%%%%%%%%%%%%%%%%%%%%%%%%%%%%%%%%%%%%%%%%%%%%%%%%%%%%%%%%%%%%%%%%%%%%%%%%%%%%%%%       
  $^{98}_{42}$Mo   &  $-5550.194(52)$    &  $-5547.8(5.5)^{\rm a}$    &     
  
                       $^{220}_{86}$Rn   &   $-27497.30(48)$          &   $-27486(170)^{\rm d}$    \\
%%%%%%%%%%%%%%%%%%%%%%%%%%%%%%%%%%%%%%%%%%%%%%%%%%%%%%%%%%%%%%%%%%%%%%%%%%%%%%%%%%%%%%%%%%%%%%%%%%%%%%%%  
                   &                     &  $-5550.5^{\rm b}$         &       
                   
                       $^{226}_{88}$Ra   &   $-29059.53(61)$          &   $-29048(200)^{\rm d}$    \\
%%%%%%%%%%%%%%%%%%%%%%%%%%%%%%%%%%%%%%%%%%%%%%%%%%%%%%%%%%%%%%%%%%%%%%%%%%%%%%%%%%%%%%%%%%%%%%%%%%%%%%%%                   
 $^{102}_{44}$Ru   &  $-6139.633(54)$    &  $-6136.8(5.8)^{\rm a}$    &       
 
                       $^{232}_{90}$Th   &   $-30687.89(60)$          &   $-30677(240)^{\rm d}$    \\
%%%%%%%%%%%%%%%%%%%%%%%%%%%%%%%%%%%%%%%%%%%%%%%%%%%%%%%%%%%%%%%%%%%%%%%%%%%%%%%%%%%%%%%%%%%%%%%%%%%%%%%% 
                   &                     &  $-6139.9^{\rm b}$         &     
                   
                       $^{238}_{92}$U    &   $-32386.00(64)$          &   $-32374(300)^{\rm d}$    \\
%%%%%%%%%%%%%%%%%%%%%%%%%%%%%%%%%%%%%%%%%%%%%%%%%%%%%%%%%%%%%%%%%%%%%%%%%%%%%%%%%%%%%%%%%%%%%%%%%%%%%%%%                   
 $^{106}_{46}$Pd   &  $-6761.805(57)$    &  $-6758.8(6.3)^{\rm a}$    &   
 
                       $^{240}_{94}$Pu   &   $-34158.6(1.1)$          &   $-34147(300)^{\rm d}$    \\
%%%%%%%%%%%%%%%%%%%%%%%%%%%%%%%%%%%%%%%%%%%%%%%%%%%%%%%%%%%%%%%%%%%%%%%%%%%%%%%%%%%%%%%%%%%%%%%%%%%%%%%% 
                   &                     &  $-6762.1^{\rm b}$         &    
                   
                       $^{244}_{96}$Cm   &   $-36009.8(1.0)$          &   $-35996(400)^{\rm d}$    \\
%%%%%%%%%%%%%%%%%%%%%%%%%%%%%%%%%%%%%%%%%%%%%%%%%%%%%%%%%%%%%%%%%%%%%%%%%%%%%%%%%%%%%%%%%%%%%%%%%%%%%%%%                   
 $^{114}_{48}$Cd   &   $-7417.323(60)$   &  $-7413.9(6.8)^{\rm a}$    &
                 
                                         &                            &                            \\
%%%%%%%%%%%%%%%%%%%%%%%%%%%%%%%%%%%%%%%%%%%%%%%%%%%%%%%%%%%%%%%%%%%%%%%%%%%%%%%%%%%%%%%%%%%%%%%%%%%%%%%% 
                   &                     &  $-7417.7^{\rm b}$         &     
                   
                                         &                            &                            \\
%%%%%%%%%%%%%%%%%%%%%%%%%%%%%%%%%%%%%%%%%%%%%%%%%%%%%%%%%%%%%%%%%%%%%%%%%%%%%%%%%%%%%%%%%%%%%%%%%%%%%%%%                                                    

\hline
\hline
\end{tabular} 
\end{center} 
\bigskip
\raggedright

$^{\rm a}$ Bi\'emont \textit{et al.} \cite{Biemont:1999:117}   
\qquad \qquad 
$^{\rm b}$ Gu \cite{Gu:2005:267}.
\qquad \qquad
$^{\rm c}$ Chung \textit{et al.} \cite{Chung:1993:1740}.

$^{\rm d}$ Rodrigues \textit{et al.} \cite{Rodrigues:2004:117} with the uncertainty 
prescribed by NIST.

$^{\rm e}$ Kramida and Reader \cite{Kramida:2006:457}.

\end{table}
}

Table~\ref{table:all} displays the ground-state binding energies of  berylliumlike ions with $Z=18-96$. For calcium, xenon,
 and uranium the averages of the values calculated using three screening potentials are shown. For other ions the results of the calculations with the LDF potential  are presented. The uncertainties, which are given in the parentheses, were 
obtained by summing quadratically  the uncertainty due to the
nuclear size effect, the uncertainty of the CI-DFS calculation, and 
the uncertainty due to uncalculated two-loop one-electron  QED contributions and uncalculated QED corrections 
of third and higher orders. For uranium the nuclear size uncertainty was estimated following to Ref.~\cite{Kozhedub:2008:032501}. For other ions this uncertainty was estimated by summing  quadratically two values. 
The first one was 
obtained by varying the root-mean-square radius within its error bar presented in Ref. \cite{Angeli:2013:69}. 
The second one accounts for the dependence of the nuclear size correction on the model of the nuclear charge distribution.
It was evaluated  as the difference of  the results obtained with
 the Fermi model and the model of homogeneously charged sphere.  
The uncertainty due to uncalculated QED corrections to the interelectronic interaction of third and higher orders was conservatively estimated multiplying the term $E^{(\geqslant3)}_{\rm int,Breit}$ by 
the double ratio of the second order QED correction to the interelectronic interaction and the  corresponding contribution calculated 
within the Breit approximation, $E^{(2)}_{\rm int,QED}/E^{(2)}_{\rm int,Breit}$. 
The uncertainty due to uncalculated 
two-loop one-electron QED terms was estimated 
following to Ref.~\cite{Yerokhin:2008:062510}. Finally, we have conservatively estimated the contribution of the  higher-order screened
 QED diagrams by multiplying the second order QED term by the factor $2/Z$. 
For low-$Z$ ions the total uncertainty is mainly  
determined by the uncertainty of the CI-DFS calculation. 
For high-$Z$ ions the uncertainties due to
the nuclear size effect and uncalculated one-electron two-loop corrections play a dominant role.
In Table~\ref{table:all} we compare our binding energies with the NIST database compilation and the relativistic CI calculations by Chen and Cheng \cite{Chen:1997:166}, Gu \cite{Gu:2005:267},
 and Yerokhin \textit{et al.} \cite{Yerokhin:2014:022509}. The main uncertainty to the NIST final values comes from the ionization potentials for berylliumlike ions. This is not surprising since, as it was mentioned above, 
all the previous
 calculations of Be-like ions include the QED effects either semiempirically or in some one-electron approximations.
 It is seen that, as a rule, 
our results are in a good agreement with the previous calculations but have much higher accuracies.
%The NIST most precise result for high-$Z$ ions is for the berylliumlike tungsten. Nevertheless, even in this case we were able to decrease the uncertainty.
Some discrepancy with the NIST values is observed for several first ions with $Z>50$. The reason of this discrepancy is 
unclear to us.

%We suppose that the reason is that after $Z=50$ NIST changes the source of the ionization potentials for berylliumlike ions. Instead of the semiempirical results of  Bi\'emont \textit{et al.} \cite{Biemont:1999:117} for $Z\leqslant 50$ NIST starts to use the results of the work by  Rodrigues \textit{et al.} \cite{Rodrigues:2004:117}.
 
Finally, in Table~\ref{table:IP} we present the ionization potentials for berylliumlike ions. 
The ionization potentials are obtained by subtracting the binding energies of Li-like ions from the binding energies 
of Be-like ions
presented in Table~\ref{table:all}. 
We do not use the NIST compilation for lithiumlike ions directly. 
Instead, we sum the ionization potentials for H-, He-, and Li-like ions. 
For heliumlike and lithiumlike ions we use the tabulations from Refs. \cite{Artemyev:2005:062104} and \cite{Sapirstein:2011:012504}, 
respectively. However, we have recalculated the finite nuclear size correction to the Dirac energies 
of the $1s$ and $2s$ states, 
since in those works the calculations were performed for nuclear radii that differ from ones we use here. 
To obtain the ionization potentials for H-like ions, we sum all the one-electron $1s$ contributions. 
The self-energy contributions 
were obtained by interpolating the values presented in Ref.~\cite{Mohr:1992:4421} with
 the nuclear size correction according to Ref.~\cite{Yerokhin:2011:012507}. The contributions of the vacuum polarization diagram 
were taken from Ref.~\cite{Sapirstein:2003:022512}. The two-loop one-electron QED corrections were added according to
 the works \cite{Yerokhin:2005:040101_R,Yerokhin:2006:253004,Yerokhin:2008:062510,Yerokhin:2009:040501_R}. To incorporate the recoil corrections we used the data presented in Ref.~\cite{Shabaev:1998:4235}.
In Table~\ref{table:IP}
our ionization potentials  for the berylliumlike ions are compared with other theoretical predictions. 
As it was mentioned above, the works \cite{Biemont:1999:117} and \cite{Kramida:2006:457}  are semiempirical. 
Therefore, one can consider the corresponding values and uncertainties as some experimental limits for 
 the ionization potentials. It can be seen that our ionization energies agree
 with the previous calculations but have a much higher accuracy, especially for high-Z ions.

\section{Summary \label{sec:3}}

To summarize, the calculations of the ground-state
binding energies and ionization potentials for berylliumlike ions were 
performed in the range: $18\leqslant Z \leqslant 96$. Our computational procedure allows to 
merge the rigorous QED calculations up to the second order of the perturbation theory with
the large-scale CI-DFS evaluations of the higher-order electron-correlation contributions.
As the result, the most precise theoretical predictions for the ground-state 
binding energies and the ionization potentials in Be-like ions have been obtained.

\section*{Acknowledgements}
This work was supported by RFBR (Grants No. 13-02-00630, No. 12-03-01140, 
No. 14-02-31316, No. 14-02-00241, and No. 14-02-31476), by SPbSU
(Grants No. 11.38.269.2014, No. 11.38.261.2014, and No. 11.0.15.2010), and by  DFG (Grant No. VO 1707/1-2).
A.V.M. acknowledges the support from the Dynasty foundation, G-RISC, and DAAD.
The work of D.A.G. was also supported by the FAIR-Russia Research Center and by
the Dynasty foundation.

%\bibliography{biblio_v3}


\end{document}